# The Impact of R&D Investments, Including AI, on Economic Growth and the Country's Capacity to Improve Its Credit Rating

**Davit Gondauri,** ORCID: https://orcid.org/0000-0002-9611-3688
Doctor of Business Administration, Business & technology University, Georgia

**Ekaterine Mikautadze,** ORCID: https://orcid.org/0009-0000-9942-1478
Doctor of Biology, European University & Alte University, Georgia

**Nino Enukidze,** ORCID: https://orcid.org/0000-0001-5067-5311
Doctor of Business Administration, Business & Technology University, Georgia

**Mikheil Batiashvili,** ORCID: https://orcid.org/0009-0004-7752-0423
Professor, Chairman of Supervisory board, Business & Technology University, Georgia

**Corresponding author:** Davit Gondauri, dgondauri@gmail.com

**Abstract:** *The research and development phase is a crucial initial step in any process leading to innovation, and it aligns with the long-term vision of public and private sector strategies. The research questions in this study are as follows: (1) To determine the resulting interrelationship between R&D investments and GDP using regression analysis; (2) To investigate the amount of economic value added (EVA) that Georgia must create with the increase of R&D investments in a certain period in order to move from the group of countries with a BB sovereign credit rating to the group of countries with a BBB-investment credit rating. World Bank data from 2014-2022 was used. Using regression analysis, the impact of R&D investments (by increasing the share of artificial intelligence in R&D to 30-35%) on the country's GDP was determined. The regression analysis between R&D and GDP generated the following results: (1) the regression coefficient is 7.02502%, indicating that a 10% increase in R&D will result in a 0.70% increase in GDP; (2) the coefficient of determination is 81.1%, which demonstrates that 81.1% of the change in GDP is explained by the change in R&D; (3) the correlation coefficient is 90.1%, indicating a strong positive relationship; (4) the P-value is 0.03492, which suggests that the relationship between these two variables is significant. The calculation of a country's EVA (as a powerful tool that evaluates a country's economic growth and development) incorporates three key factors: the country's total wealth, net operating profit after tax and the Central Bank rate. The EVA model of Georgia was calculated and then analysed in order to determine the additional value that Georgia would need to generate in order to be included in the BBB-investment credit rating group. In order to determine this, the economic indicators of the countries on the BBB scale (Greece, Hungary, India, Kazakhstan) were analysed, and their average weighted index was calculated. This index is characteristically relevant according to the criteria set out by S&P, Fitch, and Moody's. The following economic indicators were considered: nominal GDP, GDP per capita, and real GDP growth. External indicators included: current account balance/GDP, gross external financing needs/CARs plus usable reserves. Fiscal indicators were: general government balance/GDP, debt/GDP, and net debt/GDP. Finally, the consumer prices index growth was considered as a monetary indicator. According to EVA model calculations, in order to achieve Georgia's BBB credit rating in the next 9 years, investments of $61.7 billion are required. Using EVA and other economic indicators in the decision-making process will contribute to a more in-depth analysis of the current economic processes in the country and increase efficiency.*

**Received:** 17.06.2024          **Accepted:** 18.08.2024          **Published:** 02.10.2024

**Funding:** There is no funding for this research.
**Publisher:** Academic Research and Publishing UG (i.G.) (Germany).
**Founder:** Academic Research and Publishing UG (i.G.) (Germany).

**AR&P**

## INTRODUCTION

Innovation is recognized as a key tool for economic growth and increased competitiveness. Effective use of existing technologies in a country is a meaningful way to maintain economic growth and development. Research and development (R&D) plays a key role in economic growth.

Globalization accelerates the shift of trade from a local to an international dimension. At the same time, innovation is becoming a key factor for countries to increase employment, sustainability, social welfare, and quality of life. Increasing the level of R&D creates the basis for innovation (Akcali & Sismanoglu, 2015).

Regarding R&D, education and science occupy the most crucial place in human life and are insignificant for the country's development. A priority direction has no alternative. In this context, the management of R&D activities is a crucial enabler of growth, as it ultimately leads to increased knowledge, better technology, and innovative products and processes in the public and private sectors. In other words, R&D are necessary processes that bring innovation.

R&D is necessary in the first phase of any process leading to innovation. It involves the long-term vision of public and private sector strategy, while innovation focuses on the shorter term. Ultimately, R&D involves investing money to find innovative products, services, or processes that increase economic opportunity (through the innovation discovered). Accumulated experience and knowledge through R&D enhance innovation for any company (Brouwer & Kleinknecht, 1996). Investing in R&D provides the opportunity to create technology and future capabilities that can be transformed into new products, processes and services.

The study aims to investigate the possibilities of a country's transition from the BB credit rating to the BBB- credit rating. In particular, the research questions are: 1. To determine the resulting relationship between R&D-AI investments and GDP using regression analysis; 2. How much economic value added (EVA) does Georgia need to create with the increase of R&D-AI investments in a certain period in order to be able to move from the group of countries with BB sovereign credit rating to the group of countries with BBB-investment credit rating?

Funding for specialized R&D organizations in the government sector of all economies comes primarily from the government itself. The contributions of the business sector to the R&D of these organizations do not exceed 15% for any economy. In contrast, in the case of Belgium, donations from the rest of the world can be more than 40%. For several European economies, such contributions maybe 20% and are likely mainly from EU programs. Private non-profit funding is noticeable only in the case of Israel at slightly less than 10%.

The research on the mentioned issues is relevant, especially when some types of evaluations are found in the literature. In the government sector, government R&D studies should be effectively taken from the perspective of the statistical unit of the performer rather than collecting all the information from the sponsoring agencies or ministries without accidents, which include a variety of institutional and functional ranges of R&D performing organizations under their supervision. The correct interpretation of Business Enterprise Research and Development (BERD) statistics in the business sector requires a distinction between the contribution of non-profit organizations under the control of business enterprises and other profitable business enterprises and their subsidiaries. The questionnaire should be adapted to consider the different types of business enterprises.

The goods produced in Georgia in the sphere of innovation and technologies have a low benefit. For now, only the processing industry is relatively developed, which is directly connected to the natural resources obtained in Georgia and local agricultural products. The reason for this is the low level of technological development and innovation, which simultaneously leads to the irrational use of natural resources and threatens the country's natural wealth. Both state and private sector expenditures on R&D are low in the country, as reflected in various international assessments and rankings.

## LITERATURE REVIEW

Since the early 2010s, with significant advances in computer vision and speech recognition, deep learning has led to rapid artificial intelligence (AI) progress (LeCun et al., 2015; Russell and Norvig, 2020). The artificial intelligence paradigm is responsible for nearly every milestone of the past decade. Two significant trends can be used to understand its impact: (i) recent advances in deep learning in R&D and (ii) the rapid scaling of computing in deep learning systems. Artificial intelligence is based on an endogenous growth model, which means that the development of AI depends on internal innovation and technological progress, which constantly enhances its growth and development. Researchers have shown that if deep learning leads to increased capital in R&D, it can accelerate innovation and economic growth. Leveraging new computational and human capital data, if deep learning is widely adopted in the US R&D sector, will lead to an accumulation of computational capital that doubles the rate of productivity growth (Besiroglu, T., Emery-Xu, N., & Thompson, N., 2024).

The most frequently used tool for evaluating research is the productivity of science, which, on the one hand, serves to achieve several goals and is also essential for society, the economy, or science itself. Researchers and higher education institutions often use advances to evaluate proposals for research projects at the sectoral or national level. Number of publications, the primary tool for evaluation, are used to evaluate the productivity and success of a researcher or the impact of research (Aksnes et al., 2012; Kulczycki, 2017). In general, R&D funding at the national level is an important tool for establishing institutional research activities, including staff development (Henriksen & Schneider, 2014; Franzoni et al., 2012).The diversity of university management models, funding schemes, and enabling factors affects the effectiveness of publications. Tools that focus on researcher productivity positively or negatively affect the research itself (Aagaard & Schneider, 2017; Bal, 2017). Empirically measuring the relationship between innovation and future performance is a challenge. Although readable data on R&D expenditures are available, making it a natural choice as an empirical indicator of innovation effort, R&D expenditures represent only one aspect of the innovation process (Matolcsy & Wyatt, 2008).

Two key indicators of national R&D performance are gross domestic expenditure on R&D (GERD), a measure of a country's total R&D investment, and national R&D intensity (GERD-to-GDP ratio), a measure of a country's investment in R&D relative to overall economic activity. Together, they provide a broad picture of the current distribution of global R&D activities and the changing global R&D landscape as countries build opportunities in science and technology to improve their national economies and societies.

The education received gives the best results for the country's economic growth. The mentioned results of the education reform are provided by the teaching method, and sometimes this method affects R&D, for example, "10% higher R&D expenditure = 4% GDP growth". It is an example of the Countries in the Middle East and North Africa (MENA). What happens when the best teaching method allows the students in their youth to use it in the future to develop the best analytical skills, which will gradually be reflected in R&D. This does not mean that the masses should learn the best method in the same way. It is worth noting that the Pareto principle applies 20% of the effort, bringing 80% of the results, and the remaining 80% only 20%. It is by using this principle and the best teaching method that more effective R&D in the country's GDP will be obtained in the future. The remaining 80% of the Pareto principle goes into vocational education to satisfy the labor market.

Collaborative studies by Gu (2005) and Matolcsy and Wyatt (2008) extended the literature on innovation by including outcome measures based on patent citations in the research design. However, these studies still need to be answered due to their different focus. For example, Gu (2005) studied patent-related measures but did not examine inclusive measures of innovation, such as R&D expenditures, nor did he relate the variation in realized operating performance to a firm's innovative output. Matolcsy and Wyatt's (2008) study focused on the interaction between earnings and aggregate rather than firm-level measures of technological innovation. Like Gu (2005), they related earnings and technological conditions to market value rather than the volatility of future operating performance.

Artificial intelligence improves R&D capital, which, in turn, is a prerequisite for economic growth. Recent studies have shown that deep learning in R&D and the rapid increase in computing resources used by deep learning systems accelerate technological progress. Human capital valuation is 4-5 times more accurately predicted using AI than other methods used in the literature. Deep learning R&D is more capitalized than other forms, allowing for profit maximization (Besiroglu et al., 2024).

AI terms appear frequently in successful R&D applications of each funding database. The occurrence of a critical term in a particular database corresponds to the number of documents that appear at least once, normalized to the number of documents in each database for comparability. The top 10 most common terms are listed for each database. "Robot" was the most frequently used term in many databases, except those associated with medical or life science-focused agencies. "Bioinformatics" appeared more frequently in the latter databases. Neither of these terms are considered primary key AI terms, as they can also be used in AI-related contexts (Yamashita et al., 2021).

The EVA model has qualitatively changed the corporate management and evaluation system at the modern stage. Researchers in the EVA model attach fundamental importance to R&D as a strategic capital cost. At the modern stage, with EVA modelling, it is possible to analyze the economic effect of R&D investments, which can influence the management of the R&D portfolio and the generation of technical ideas that present the role of R&D as an investment in the development of the corporation (Hatfield, 2002).

**AR&P**

## METHODOLOGY

The research mainly includes data collection, analysis, and calculations using quantitative methods. The analysis structure consists of three stages:

1. Using regression analysis, to determine the causal relationship between R&D-AI and GDP. Regression analysis is an effective method for determining how strongly and in what way R&D and AI investments are related to the growth of a country's GDP. This research is quantitative and based on statistical methods.

2. To develop and analyze the country's economic growth and development model using EVA.

3. After calculating the EVA, the next step is to analyze the sovereign credit rating (BB scale) of Georgia and compare it with the average rate of countries with a BBB rating. After that, to calculate and analyze what added value to create and what period it would take for Georgia (due to the increase in R&D-AI investments) to move to the ranks of BBB-rated countries.

World Bank data from 2014-2022 was used to calculate the regression analysis between R&D-AI and GDP. At the next stage, the mentioned data were logarithmized (with Neper's base), based on which the regression coefficient was calculated in Equation 1.

$$b_1 = \frac{\sum_{i=1}^{n}(x_i - \bar{x})(y_i - \bar{y})}{\sum_{i=1}^{n}(x_i - \bar{x})^2}, \tag{1}$$

where $b_1$ is the slope of the regression line in simple linear regression. It represents the change in the dependent variable $y$ for a one-unit change in the independent variable $x$; $x_i$ is the $i$-th value of the independent variable $x$. It represents one data point in the dataset; $\bar{x}$ is the mean (average) of all the $x$ values in the dataset; $y_i$ is the $i$-th value of the dependent variable $y$. It represents one data point in the dataset; $\bar{y}$ is the mean (average) of all the $y$ values in the dataset.

After the regression coefficient, the Pearson correlation coefficient is calculated in equation 2.

$$r = \frac{\sum(x - \bar{x})(y - \bar{y})}{\sqrt{\sum(x - \bar{x})^2 \sum(y - \bar{y})^2}}, \tag{2}$$

where $r$ is the Pearson correlation coefficient. It measures the strength and direction of the linear relationship between two variables x and y; $x$ represents a specific value of the independent variable; $\bar{x}$ is the mean (average) of all the x values in the dataset; y represents a specific value of the dependent variable; $\bar{y}$ is the mean (average) of all the y values in the dataset and the coefficient of determination – $(3)=(2)^2$

And finally, to test the correctness of the hypothesis in equation 3.

$$P = 2 \cdot P(T > |t|), \tag{3}$$

where T is the t-distribution, and t is the t-statistic, $P$ is the Pearson correlation coefficient.

The application of EVA for a country is more complex and multifaceted than for a company, however, it can be used to analyze the economic growth and development of a country.

*Calculation of EVA for a country*

Step 1: Determine the NOPAT across the country

Net Operating Profit After Taxes (NOPAT) can be defined as GDP of a country after taxes in equation 4.

$$NOPAT = GDP \cdot (1 - ATR), \tag{4}$$

where *NOPAT* is Net Operating Profit After Taxes, *GDP* is Gross Domestic Product, *ATR* is Aggregate Tax Rate.

Step 2: Determination of the country's Total Wealth (*TW*)

*TW* of a country includes the sum of manufactured, human, natural capital and net foreign assets used in the economy is calculated in equation 5.

$$TW = PC + HC + NC + NFA, \tag{5}$$

where, *PC* is the Produced capital, *HC* is the Human capital, *NC* is the Natural capital, *NFA* is the Net foreign assets.

Step 3: Determination of the Central Bank Rate (*CBR*).

It is an instrument of monetary policy and includes the equilibrium value added to various financial instruments and investments. The weighted average of the last 10 years is taken

Step 4: Calculate *EVA*

*EVA* is the difference between N*OPAT* and cost of capital is presented in equation 6.

$$EVA = NOPAT - (TW \cdot (CBR)), \tag{6}$$

where *EVA* is Economic Value Added, *NOPAT* is Net Operating Profit After Taxes, *CBR* is the Central Bank Rate, *TW* is the Total wealth.

After calculating the EVA model of Georgia, the evaluation criteria of S&P, Fitch, Moody's sovereign credit ratings were searched and analyzed from the websites of credit agencies (S&P Global, 2024; Fitch Ratings, 2024; Moody's Investors Service, 2024). Specifically, it was analyzed how much additional value Georgia needs to generate to get into the BBB-investment credit rating group. /To determine this, at the current stage, the economic indicators of the countries on the BBB scale (Greece, Hungary, India, Kazakhstan) were analyzed, and their average weighted index was calculated, which is characterized by high relevance according to S&P, Fitch, Moody's criteria, namely:

Economic indicators (%): nominal GDP (bil. $), GDP per capita (000s $), real GDP growth.

External indicators (%): current account balance/GDP, gross external financing needs/CARs plus usable reserves.

Fiscal indicators (general government; %): Balance/GDP, Debt/GDP, Net debt/GDP.

Monetary indicators (%): the Consumer Prices Index (CPI) growth.

**RESULTS**

Research is complex in many ways. On the one hand, the question concerns the impact of R&D-AI investments on the country's GDP, using regression analysis.

Then it is calculated on the example of the country how much-added value is created in the GDP by R&D-AI investments, and in the third stage, it is analyzed how much-added value is needed and for what period for Georgia to get its sovereign credit rating (according to S&P, Fitch, Moody's) which is currently in the credit ratings. In the non-investment group, to improve from the BB scale and move to the investment group on the BBB scale.

Table 1 presents all the economic indicators that were used to calculate the EVA created by Georgia in 2014-2022.

**Table 1. Economic Indicators of Georgia in 2014-2022**

| Georgia (bln $) | 2014 | 2015 | 2016 | 2017 | 2018 | 2019 | 2020 | 2021 | 2022 |
|---|---|---|---|---|---|---|---|---|---|
| GDP | 17.63 | 14.95 | 15.14 | 16.24 | 17.60 | 17.47 | 15.84 | 18.63 | 24.78 |
| Aggregate tax rate, % | 12.2 | 12.2 | 12.2 | 12.2 | 12.2 | 12.2 | 12.2 | 12.2 | 12.2 |
| NOPAT=GDP·(1−Aggregate tax rate) | 15.48 | 13.13 | 13.30 | 14.26 | 15.45 | 15.34 | 13.91 | 16.36 | 21.76 |
| Central Bank Rate (weighted average), % | 8.3 | 8.3 | 8.3 | 8.3 | 8.3 | 8.3 | 8.3 | 8.3 | 8.3 |
| Produced capital | 74.58 | 76.79 | 79.35 | 81.86 | 84.56 | 86.11 | 87.68 | 89.29 | 90.93 |
| Human capital | 61.80 | 61.87 | 63.70 | 64.23 | 67.35 | 68.59 | 69.84 | 71.12 | 72.43 |
| Natural capital | 14.32 | 14.72 | 13.96 | 14.54 | 14.55 | 14.82 | 15.09 | 15.37 | 15.65 |
| Net foreign assets | (15.29) | (19.28) | (21.22) | (23.30) | (22.95) | (22.61) | (22.28) | (21.95) | (21.62) |
| Total wealth | 135.41 | 134.10 | 135.80 | 137.33 | 143.51 | 146.90 | 150.34 | 153.84 | 157.38 |
| **EVA = NOPAT - Total wealth · Central Bank Rate** | **4.31** | **2.07** | **2.09** | **2.93** | **3.61** | **3.22** | **1.51** | **3.67** | **8.78** |

*Source: compiled by the authors.*
*Notes: GDP - Gross Domestic Product, EVA - Economic Value Added, NOPAT - Net Operating Profit After Taxes.*

**AR&P**

In the mentioned years, the average annual EVA amounted to 3.58 billion, the minimum EVA was 1.51 billion in 2020, and the maximum was 8.78 billion in 2022. Stable growth of R&D-AI investments is of great importance for the development of Georgia's economy, EVA, and hence, the country's sovereign credit rating and the transition to the country's investment group because R&D increases both technological and innovative progress, as well as improves the country's education level.

Figure 1 depicts the research and development expenditure (% of GDP) in 2013-2022 according to the countries of the region.

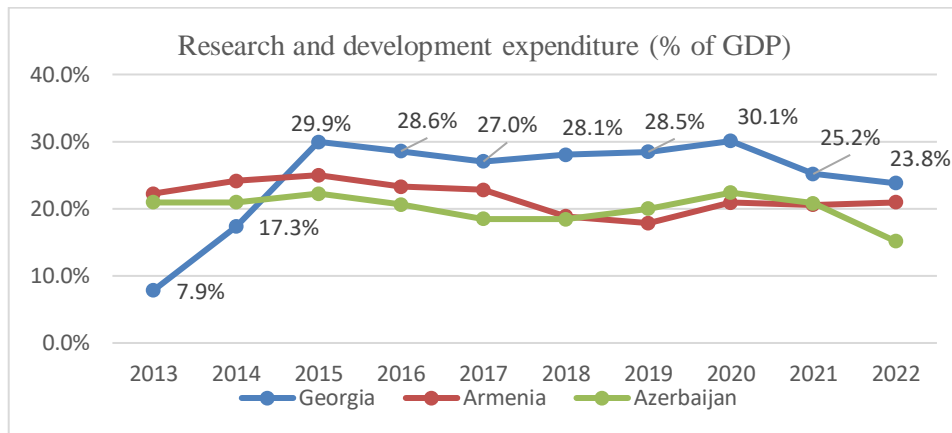

**Figure 1. Research and Development Expenditure (% of GDP) in 2013-2022 is Presented according to the Countries of the Region**

*Source: compiled by the authors.*
*Notes: GDP - Gross Domestic Product.*

Georgia is undoubtedly the leader in the region. The average annual share of research and development expenditure (% of GDP) in 2013-2022 was 24.6%, Armenia - 21.7%, and Azerbaijan - 20%.

Globally, in 2013-2022, according to Research and development expenditure (% of GDP), the five leaders are Israel - 5.56%, South Korea - 4.93%, USA - 3.46%, Belgium - 3.43%, Sweden - 3.42%.

Georgia occupies 56 places in the mentioned list. Along with many other economic indicators and factors, the lag in Research and development expenditure is one of the main reasons the BB level of Georgia's sovereign credit rating is still in the non-investment group. Depending on the purpose of the study, a regression analysis was also conducted between R&D and GDP. As a result of the Research, it was established that the regression coefficient is 7.02502% (multiplier); the coefficient of determination is 81.1%; 3. the correlation coefficient is 90.1%; the P-value is 0.03492. The regression coefficient is 7.02502%, which means that if R&D increases by 10%, then GDP will increase by 0.70%. The correlation coefficient (90.1%) indicates a strong positive relationship between the two variables. The coefficient of determination (81.1%) suggests that it explains the relationship between R&D and GDP by 81.1%. The P-value (0.03492) means that it is less than 0.05, and based on this. Hence, the null hypothesis is rejected, and the relationship between these two variables is significant. In the next stage of the study, the resulting multiplier was an average increase in R&D of 10% annually over the next 9 years, followed by a rise in GDP of 0.70%.

Based on this, the corrected EVA model of Georgia for the next 9 years will take the following form, according to Table 2.

**Table 2. The Corrected EVA Model of Georgia for the Next 9 Years**

| GEORGIA (bln $) | 1 | 2 | 3 | 4 | 5 | 6 | 7 | 8 | 9 |
|---|---|---|---|---|---|---|---|---|---|
| GDP | 24.95 | 25.13 | 25.31 | 25.48 | 25.66 | 25.84 | 26.03 | 26.21 | 26.39 |
| Aggregate tax rate, % | 12.2 | 12.2 | 12.2 | 12.2 | 12.2 | 12.2 | 12.2 | 12.2 | 12.2 |
| NOPAT=GDP · (1−Aggregate tax rate) | 21.91 | 22.07 | 22.22 | 22.38 | 22.53 | 22.69 | 22.85 | 23.01 | 23.17 |
| Central Bank Rate (weighted average), % | 8.3 | 8.3 | 8.3 | 8.3 | 8.3 | 8.3 | 8.3 | 8.3 | 8.3 |
| Total wealth | 135.48 | 134.17 | 135.88 | 137.42 | 143.60 | 147.00 | 150.46 | 153.96 | 157.52 |
| **EVA = NOPAT - Total wealth · Central Bank Rate** | **10.74** | **11.00** | **11.01** | **11.04** | **10.69** | **10.56** | **10.44** | **10.31** | **10.18** |

*Source: compiled by the authors.*
*Notes: GDP - Gross Domestic Product, EVA - Economic Value Added, NOPAT - Net Operating Profit After Taxes.*

In the next 9 years, according to the presented model, the average annual EVA of Georgia will be 10.66 billion, the minimum EVA – 10.18 billion in the ninth year, and the maximum – 11.04 billion in the fourth year. The total present value (PV) of EVA created in 9 years is 66.2 billion dollars.

Figure 2 shows that as a result of investments in Research and development expenditure, the share of R&D in GDP will increase from 0.23% to 0.53% in 9 years if R&D rises by 10% per year on average.

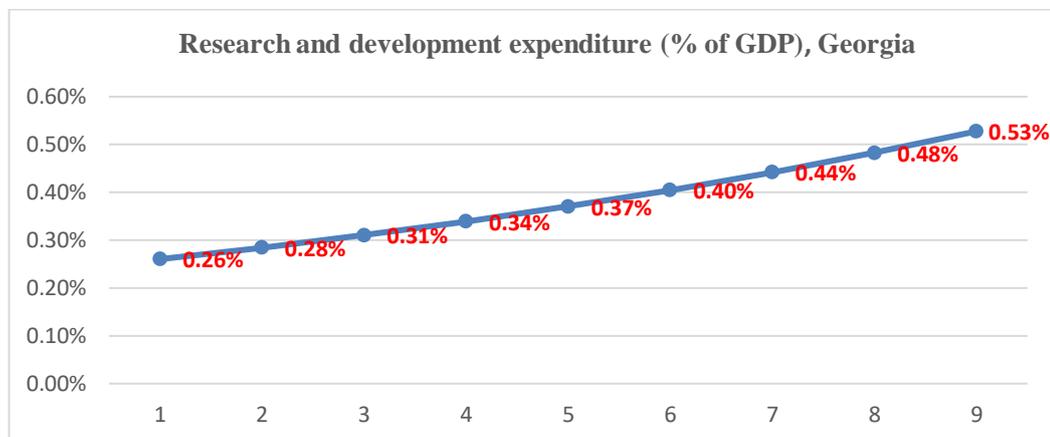

**Figure 2. Research and Development Expenditure (% of GDP) for 9 years**

*Source: compiled by the authors.*
*Notes: GDP - Gross Domestic Product.*

If Georgia increases the share of AI in R&D to 30-35%, with the multiplier effect, R&D will already be able to increase GDP with the effect presented in the regression analysis. For the future effectiveness of the EVA model, Georgia must ensure the protection of the mentioned proportion of AI in R&D.

After this part of the study, our goal is to analyze the sovereign credit rating of Georgia and to show what added value the country needs to move from BB to BBB, to finance negative economic processes, and how many years Georgia will be able to advance.

Table 3 presents S&P's leading economic, foreign, financial and monetary indicators strongly influencing country ratings.

**Table 3. S&P's Main Economic, Foreign, Financial and Monetary Indicators**

| 2024 | BB | BBB- | BBB- | BBB- | BBB- | BBB - Average |
|---|---|---|---|---|---|---|
| | Georgia | Greece | Hungary | India | Kazakhstan | |
| **ECONOMIC INDICATORS (%)** | | | | | | |
| Nominal GDP (bil. $) | 30 | 255 | 222.8 | 3,909 | 298 | 258.6 |
| GDP per capita (000s $) | 8.1 | 24.5 | 23.3 | 2.7 | 14.7 | 20.83333333 |
| Real GDP growth | 4.5 | 2.4 | 2.2 | 6.8 | 3.5 | 3.725 |
| **EXTERNAL INDICATORS (%)** | | | | | | |
| Current account balance/GDP | -4.7 | -5.7 | 0.4 | -1.3 | -2.8 | -2.35 |
| Gross external financing needs/CARs plus usable reserves | 113.7 | 314.1 | 107.5 | 82.7 | 99.2 | 96.46666667 |
| **FISCAL INDICATORS (GENERAL GOVERNMENT; %)** | | | | | | |
| Balance/GDP | -2.5 | -0.5 | -5.3 | -7.9 | -2.7 | -4.1 |
| Debt/GDP | 42.8 | 150.7 | 75.6 | 84.3 | 26.1 | 84.175 |
| Net debt/GDP | 38 | 136.3 | 72.3 | 84.3 | 3.6 | 74.125 |
| **MONETARY INDICATORS (%)** | | | | | | |
| CPI growth | 3.2 | 2.8 | 4.4 | 4.5 | 8 | 4.925 |

*Source: compiled by the authors.*
*Notes: GDP - Gross Domestic Product, CPI - the Consumer Prices Index growth.*

In this case, the study's goal is to move Georgia to the investment scale of S&P, which starts from BBB. Georgia is now on BB.

Countries are presented in the table next to Georgia, and Robles are also on BBB- rating. Their weighted average is shown in the right corner; their average was compared with Georgia's data.

According to the data presented in green color in the average column, Georgia is better than the average of other countries. What is brown? The average of the data is better than the indicator of Georgia.

Based on Table 3, the following financial resources and investments are needed to achieve a BBB credit rating for Georgia:

Nominal GDP requires about $228.6 billion of investments in the economy, although it is possible to compensate with the rest of the indicators;

GDP per Capita will require investments of $47.7 billion;

Real GDP Growth. The country should maintain a growth rate of 4.5%, which is better than the BBB-average of 3.725%. For this, the European financial resources require an investment of about $5 billion annually;

Current Account Balance/GDP. Reducing the current account deficit from -4.7% to an average of -2.35% requires an increase in exports and a decrease in imports, which requires an investment of about $3 billion in improving trade relations;

Gross External Financing Needs/CARs Plus Usable Reserves Objective: to reduce the ratio from 113.7% to an average of 96.47%. An investment of approximately $5 billion will be required to mitigate external financing needs.

CPI Growth to control the level of inflation, it is necessary to tighten the monetary policy, which requires an investment of about $1 billion.

From this, by omitting the first point and summing up the others, the necessary investment of 47.7 + 5 + 3 + 5 + 1 = 61.7 billion dollars is obtained to finance the damaging processes of Georgia, which can be generated in 9 years presented by the adjusted EVA model according to Table 2.

**CONCLUSIONS**

Stimulating Georgia's economic growth and improving its credit rating is directly related to R&D investments, which are one of the key factors in the country's competitiveness in today's global economy. The analysis shows that increasing R&D investment not only enhances economic potential, but also creates long-term stability and growth opportunities. A regression analysis revealed within the study showed a clear and strong correlation between R&D investments and gross domestic product (GDP) growth. In particular, it was determined that a 10% increase in R&D investments increases GDP by about 0.70%, which directly indicates the importance of innovative technologies and research activities for Georgia's economic progress.

The mentioned results emphasize that it is necessary to apply additional resources in the field of research and development. For developing countries like Georgia, the development of innovation and technology becomes critically important not only for short-term but long-term economic sustainability. R&D investments contribute to the creation of new jobs, the introduction of innovations and the growth of competitiveness at the global level, which is directly related to the economic prosperity of the country.

For the Government of Georgia, improving the country's credit rating is one of its strategic goals, as it reflects on the country's sovereign finances, the confidence of international investors, and the terms of access to credit. As a result of the economic analysis performed within the framework of the study, it was revealed that achieving BBB credit rating of Georgia requires significantly increased investments in R&D and economic values. In order to achieve this goal, it is important that economic policy makers make decisions not only within the framework of the EVA (Economic Value Added) model, but also based on other important economic indicators, such as fiscal indicators, foreign liabilities, current account balance and inflation indicators.

*Recommendations based on research*

- Systematic use of EVA: To analyze the economic development of the country, regular calculation and monitoring of EVA should become a practice. This will help to assess and manage the efficiency of the current economic activity in the country.
- Improving data quality: The data used to determine Invested Capital and Total Wealth should be reliable and up-to-date. A full and comprehensive assessment of the country's assets will be required.
- Alternative methods of estimating cost of capital: The use of WACC, Central Bank Rate (CBR) and YTM of bonds should be considered and selected according to the situation. It is important to select them according to the country's economic specifics and market conditions.

- Strategic development of R&D and AI investments: The country should develop and increase investments in R&D and artificial intelligence (AI). These investments will help accelerate economic growth and increase the country's competitiveness.
- Credit rating improvement strategy: A strategic plan for creating additional value is needed to move Georgia's sovereign credit rating to the BBB level. Investments in sectors that provide high income growth and economic stability should be evaluated and implemented.
- Support for decision-making: the use of EVA and other economic indicators in the decision-making process will contribute to a more in-depth analysis of the current economic processes in the country and increase efficiency. This ensures the achievement of long-term and sustainable economic development.

**Author Contributions:** Conceptualisation: D. G. and E. M.; data curation: E. M.; formal analysis: D. G.; investigation: D. G. and E. M. and N. E.; methodology: D. G.; project administration: D. G. and M. B.; supervision: D. G. and M. B.; validation: M. B. and N. E.; visualisation: E. M.; writing – original draft: E. M. and N. E.; writing – review & editing: D. G. and M. B.

**Acknowledgements**

Not applicable.

**Conflicts of Interest**

Authors declare no conflict of interest.

**Data Availability Statement**

Not applicable.

**Informed Consent Statement**

Not applicable.

**AR&P**